\documentclass[10pt]{article}
\usepackage[T2A]{fontenc}
\usepackage[russian, english]{babel}
\usepackage[table,usenames,dvipsnames]{xcolor}
\usepackage{graphicx}
\usepackage{wrapfig}

\usepackage[backend=biber, style=authoryear, useprefix=false, maxcitenames=2, maxbibnames=30, dashed=true, natbib]{biblatex}

\usepackage{comment}
\usepackage{multirow}
\usepackage{placeins}
\usepackage{csquotes}
\usepackage{float}
\usepackage[font=scriptsize,labelfont=bf]{caption} 
\usepackage{amsmath}
\usepackage{amsfonts}
\usepackage{amssymb}

\definecolor{navyblue}{rgb}{0.0, 0.0, 0.5} 
\definecolor{my_red}{rgb}{1.0, 0.08, 0.0} 
\usepackage[colorlinks=true, citecolor = navyblue,  linkcolor=navyblue, filecolor=magenta, urlcolor=navyblue]{hyperref}

\usepackage{kbordermatrix}

\newcommand{\p}[1]{{\mathbb{P}} \left( \, #1 \, \right) }

\providecommand{\keywords}[1]
{
  \small	
  \textbf{\textit{Keywords---}} #1
}

\definecolor{light-gray}{gray}{0.9}
\definecolor{light-gray2}{gray}{1}

\title{Basic model for ranking microfinance institutions}
\date{}

\makeatletter
\def\@fnsymbol#1{\ensuremath{\ifcase#1\or *\or \dagger\or \ddagger\or \mathparagraph\or \mathsection\or \|\or **\or \dagger\dagger
   \or \ddagger\ddagger \else\@ctrerr\fi}}
\makeatother

\author{ Dmitry Dudukalov$^{1}$,
Evgeny Prokopenko$^{1}$, \and 
Ekaterina Savinkina$^{1,}$\thanks{{\footnotesize Sadly, Ekaterina Savinkina passed away in March $29th, 2024.$  The most of the work were done with her. The best memories to our dear friend.}} \\[1ex]
{\small $^{1}$Sobolev Institute of Mathematics,  Novosibirsk, Russia;}\\
{\footnotesize d.v.dudukalov@math.nsc.ru}\\
{\footnotesize prokopenko@math.nsc.ru}}

\begin{document}

\maketitle
\begin{abstract}
    This paper discusses the challenges encountered in building a ranking model for aggregator site products, using the example of ranking microfinance institutions (MFIs) based on post-click conversion. We suggest which features of MFIs should be considered, and using an algorithm based on Markov chains, we demonstrate the ``usefulness'' of these features on real data. The ideas developed in this work can be applied to aggregator websites in microinsurance, especially when personal data is unavailable. Since we did not find similar datasets in the public domain, we are publishing our dataset with a detailed description of its attributes.
\end{abstract}

\noindent\keywords{ ranking, microfinance, post-click conversion.}

\vskip 2pt


\section{Introduction} The microloan market had its origins in Bangladesh in the 1980s when the Nobel Peace Prize-winning economist Muhammad Yunus founded Grameen Bank to lend small amounts of cash to poor local people, almost always women, who could use it to found or sustain a small business. The success of the Grameen Bank has inspired many imitators, and encouraged commercial banks in many developing countries to take up microcredit lending as well (\citet{Yunus}).

Nowadays around $90$\% of persons in developing nations still do not have access to financial institutions for loans or savings. Therefore, microfinance is viewed as a critical tool for breaking the vicious cycle of poverty, which is characterised by low income, low savings, and low investment. In terms of development and social effect, the microfinance business enhances the quality of life for microentrepreneurs in the world’s less developed nations. Finally, having access to banks and improved security develops a sense of enterprise, which increases self-esteem and reputation. And since forecasts show that just $5$\% of microcredit demand is now being met, the microfinance industry is expected to overgrow in the coming years (\citet{almaamari}).

Initially characteristic of only developing regions, today MFIs also provide their services in developed countries. Even though the poverty rates in developed countries are moderate compared with those in developing countries, in Europe and USA there is a relevant segment of the total population without access to financial services, and a growing segment of the population is at risk of poverty. Authors of \citet{pedrini} provided a comparison of models for microfinance in developed and developing countries.

\subsection{Microloans in Russia}
History of microfinance institutions in Russia can be traced back to early $90$s when the first credit cooperatives appeared. Such organizations were necessary to finance the incipient small business with still undeveloped banking system. However, at this stage, there was no regulatory framework for MFIs, where the concepts of ``microfinance'' and ``microfinance activity'' would be defined. Later, certain areas of microfinance started new types of lending in the form of consumer credit cooperatives (including agricultural ones) and mutual lending societies. From $2004$ to $2010$, the number of consumers of microfinance services increased by more than $5$ times. Since $2011$, the activities of MFIs in Russia have been regulated by the law ``On microfinance activities and microfinance organizations''. From $2012$ to $2018$, due to the enhanced supervision of the Bank of Russia, the number of MFIs entitled to carry out banking operations decreased by $20.1$\%. But considering that the predominantly weak players left the microfinance market, the reduction in the number of institutions does not affect the development of the segment as a whole: the total portfolio of transactions of MFIs in the country is increasing. If we take the 3rd quarter of $2019$ as the base period, then in the same period in $2020$ the growth rate is $12.43$\%, and in $2021$ -- $51.74$\%. It also worth noting that the amount of microloans issued to individuals is constantly increasing (\citet{tsvetkov,  Trushina2021}). 

The COVID-19 pandemic obviously played big role in the development of microfinance institutions. The number of unemployed in Russia in $2020$ amounted to about $4.3$ million people, which is $24.7$\% higher than the previous year. It is worth considering the fact that during this period of time there was a decrease in the level of income of citizens. So for many people, microfinance has become a way out of this situation (\citet{nikolenko}).

The distinguishing traits of microloans in Russia are (\citet{Trushina2021}): borrowing a small amount of money;  short period of time; high interest rates; ease in obtaining (most MFIs only require a passport).
Also Russian microfinance institutions provide loans intended not only for the development of small businesses but for personal needs: daily living expenses, consumer goods, and paying off other loans.

The activities of Russian MFI  have a dual evaluation: on one side, they offer loans to clients who are unable to secure them from other financial sectors, making them widely and easily accessible. However, on the other side, these loans come with extremely high interest rates, often lack thorough creditworthiness assessments, and sometimes involve illegal methods for debt collection (\citet{Shaker2019})
(similar situation exists in other countries too, see \citet{Bloomberg} for stories from Cambodia, Jordan, Sri Lanka, and Mexico).
Then why should we rank them at all? First, if a person needs money and is too poor to qualify for a bank loan, they will eventually turn to microloans. Second, we can deal only with MFIs whose activities are controlled by the government. Third, ubiquitous digitalization and -- once again -- the COVID-19 pandemic urged many MFIs to move online (\citet{nikolenko}, \citet{cbr}) which makes it possible to gather information about them on one aggregator website. Having such information will help borrowers to compare different MFIs and choose the one they like most. And this is where we can help them with our ranking model.

Ideally, microfinance institutions should strike a proper equilibrium between a social or developmental objective and achieving financial targets simultaneously (\citet{Christen2000}). However, when microfinance institutions (MFIs) aim to become financially self-sustainable, they often prioritize making profits. This can lead to concerns that they might lose sight of their main goal, which is to help the poorest borrowers improve their financial situation and escape poverty (\citet{CullMorduch2017}). As a result, evaluating the performance of MFIs becomes crucial in understanding how effective they are in reducing poverty through microfinance (\citet{Hermes2018}). From this perspective, our study introduces a new approach to evaluating the performance of MFIs using conversion data, which potentially sheds light on what \citet{Shabiha2021} referred to as a blurred area of research.

There are modelling studies of various MFI characteristics in the literature. For instance, the decision of the Bank of Russia to exclude an MFI from the register list is predicted in \citet{Beketnova2020} to automate the detection of unscrupulous MFIs. In \citet{HERMES2011938}, total expenses are modelled to investigate whether there is a trade-off between outreach to the poor and the efficiency of MFIs. In \citet{MERSLAND2009662}, two types of financial performance (Return on Assets, Operational Self-Sufficiency, Portfolio Yield, and Operational Costs) and outreach performance (Average Loan Size and the Number of Credit Clients) are modelled to assess the impact of MFI governance on these indicators. These three studies use global MFI attributes as independent variables, such as authorised capital at the end of the year, number of founders (individuals and legal entities), gross loan portfolio, percentage of female borrowers, female CEO, firm size, etc. We use data on individual transactions, which better reflect the current state of affairs of MFIs and, due to the large amount of data, require a mathematical algorithm.

\subsection{Problem description}

Suppose there is an aggregator website and a list of MFIs that have signed a contract with the website. According to the contract, if a client takes a microloan in one of the MFIs from the list, using the website, this MFI pays the aggregator website a certain amount of money. 
The rank order in which MFIs are displayed on the website has a significant impact on the distribution of clients to MFIs (see Section~\ref{sec_data}, where we describe the real data in details) and, consequently, on the subsequent approval of microloans for clients. 
The aim of this paper is to present baseline algorithm for ranking MFIs that has good performance, fast calculation speed, is easy to interpret, and is not based on expert opinion. Ideally, we should be able to explain why our ranking algorithm puts one MFI higher than another.
We can use such a baseline algorithm to measure the quality of new algorithms, for example, those aimed at maximizing profit or approval rate.
The last is not an easy problem, since  increasing fees significantly does not guarantee higher profitability, and the advantages of reducing costs decrease when catering to wealthier clients ( mentioned by \citet{Cull2007}).

It's also worth noting that in this context there are fundamental limitations to building a system of personalised recommendations that works well for music, movies and so on (\citet{Aggarwal16}).
The main reason for this being that borrowers are reluctant to share their personal data. We cannot implement сollaborative filtering or сontent-based approach because there is no useful notion of user or item similarity in this context and our study shows that the available personal information, e.g., operating system, does not contribute significantly to the construction of a ranked list of MFIs (see Remark in the end of Section~\ref{sec_conversion_dataset}). For example, having an iPhone may theoretically increase the probability of getting a loan, but it seems too far-fetched to assume that one MFI will approve applications submitted by iPhone owners more often than another MFI. Therefore, we settled on building a global ranking model.

Based on the data analysis, we formed several key features which we then used to build our baseline ranking algorithm (Section~\ref{rank_alg}). 
We applied A/B testing and our own comparison method (Section~\ref{sec_comp_method}) to evaluate the algorithm's performance. During our research, we faced a number of challenges that deserve detailed discussion. Brief analysis of them as well as some ideas concerning further development of ranking algorithm can be found in Section~\ref{sec_discus}. All datasets are available on github (\citet{github}) and  some helpful supplements are in Appendix (Section~\ref{sec_appendix}).

\section{Microloans Data}\label{sec_data}

The data for building the ranking model were obtained from the aggregator website (AW) https://vsezaimyonline.ru, which contains information on various MFIs operating in Russia. All data can be divided into three parts:
\begin{itemize}
     \item Conversion dataset. Each row includes $21$ attributes that describe the completed loan application 
     and identify the user.
     \item Product dataset  provides information about each MFI.
     \item Click dataset. An extended version of the conversion dataset that includes applications with no conversions.
\end{itemize}

To better understand the data we visualize several main characteristics for all MFIs if it is feasible, and for $6$ MFIs, which we will call \textit{indicative MFIs}, otherwise. Even such a small set of MFIs is enough to show significant difference in MFIs.

\subsection{Conversion dataset} \label{sec_conversion_dataset}

\begin{wrapfigure}{H}{0.2\textwidth}
    \centering
\includegraphics[width=0.2\textwidth]{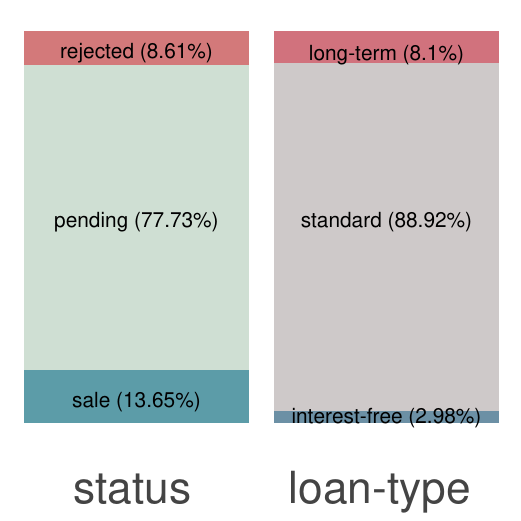}
\caption{
\textit{Status} and \textit{loan type} distributions.}
\label{status_and_loan-type}
\end{wrapfigure}

 For each application we have the id of MFI the client applied to (\textit{MFI id}) and the type of loan (\textit{loan type}). Loans can be standard, long-term, and interest-free with latter being suggested by some MFIs to attract new clients (see Fig.~\ref{status_and_loan-type}). For each pair (MFI, loan type) a unique identifier (\textit{card id}) is assigned and we will call such a pair ``MFI card'' for short.

When searching for an MFI, the clients use filters (e.g., ``Long-term loans'', ``transferring money to a bank card'', ``with a monthly payment'') which gets them to the page (\textit{page id}) with a list of suitable companies. The order of MFIs in this list is determined by some ranking method with \textit{MFI page rank} indicating MFI's position. For technical reasons, this parameter is not always defined. 
The ranking is done using the \textit{global rank} attribute in the following way: once every two weeks the AW, based on its expert's opinion, creates a ranked list of all MFI cards; this list shows \textit{MFI global rank}. After words for each page the list is filtered according to this page conditions, thus giving the \textit{MFI page rank}.

For each application, we know the moment of clicking on the MFI card (\textit{click time}), the moment of completing the application on the MFI website (\textit{conversion time}), and the moment of transferring the loan to the client if approved (\textit{sale time}). All times are in the same time zone (GMT$+7$).

After sending the application to the chosen MFI, it is assigned the status ``pending'' that later changes to ``sale'', if the application is approved, or ``rejected'' otherwise (\textit{status}). This status is relevant at the time of data upload. If the application is approved, then the MFI pays the AW a certain amount of money (\textit{income}).

When visiting the AW for the first time, each client is assigned a unique number (\textit{client id}), which does not change in the future unless the client changes the device. Using client id, we can identify different applications submitted by the same client. The attributes that identify the client are: geographic location during the session (\textit{country}, \textit{region}, \textit{city}), characteristics of the device (\textit{device type}, \textit{device}, \textit{OS}, \textit{device browser}), and type of connection (\textit{connection type}, \textit{device provider}). For convenience all the conversion dataset attributes are listed in Table~\ref{conversion-table} in Appendix.

The conversion dataset contains $67$ MFIs and $137286$ clients with $173784$ applications, $23730$ of which were approved (recall that approved applications are also called ``sales'').

\begin{table}[!ht]
\small
\begin{tabular}{||p{1.2cm}||p{0.7cm}p{0.9cm}p{0.6cm}p{1cm}p{1cm}p{0.8cm}p{0.7cm}p{0.7cm}||}
\hline
\textbf{MFI id} & \textbf{num. of app.} & \textbf{num. of unique clients} & \textbf{num. of sales} & \textbf{Income, mean} &\textbf{Income, $\sigma$} & \% \textbf{all app.} & \textbf{\% sales} & \textbf{\% all sales} \\ \hline
MFI 20 & 14635 & 14581 & 762 & 114.2 & 0.1 & 8.4 & 5.2 & 3.2 \\
MFI 87 & 11201 & 10992 & 671 & 99 & 14.5 & 6.4 & 6 & 2.8 \\
MFI 56 & 8302 & 8219 & 401 & 33.3 & 56.1 & 4.8 & 4.8 & 1.7 \\
MFI 66 & 8293 & 8288 & 609 & 231.5 & 15 & 4.8 & 7.3 & 2.6 \\
MFI 18 & 5503 & 5485 & 1355 & 175.5 & 54.8 & 3.2 & 24.6 & 5.7 \\ 
\hline
\rowcolor{white}
sum & 47934 &  & 3798 &  &  & 27.6 &  & 16 \\ 
\hline
\end{tabular}
\caption{\label{stat-table} Statistics for large MFIs shows the difference of MFIs in the percentage of sales, number of applications and income for the AW.}
\end{table}
\FloatBarrier

\begin{figure}[H]
\includegraphics[width=10cm]{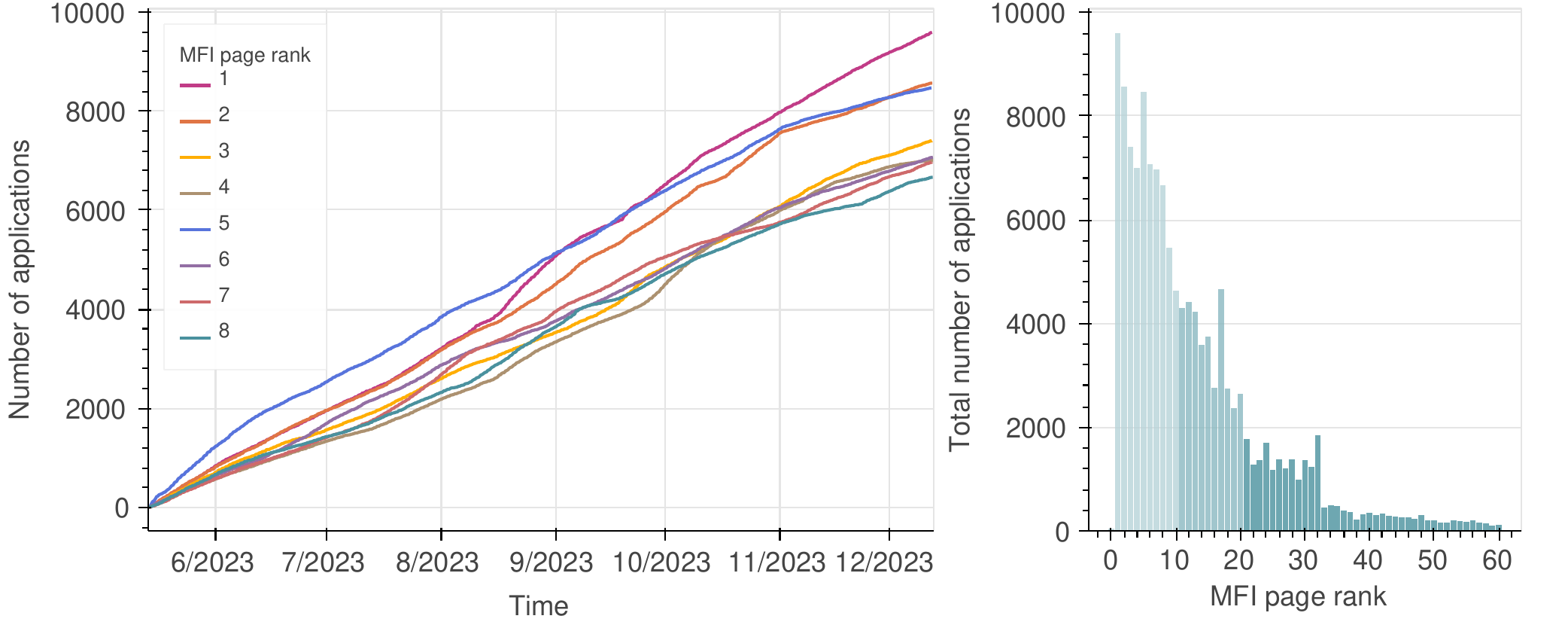}
\centering
\caption{ Dependence of the number of applications on MFI page rank. Left: cumulative number of applications increases with time (for eight highest ranks); right: the number of applications at the last available timestamp. }
\label{Number_requests_for_rank}
\end{figure}

\begin{figure}[H]
\includegraphics[height = 0.35\textwidth]{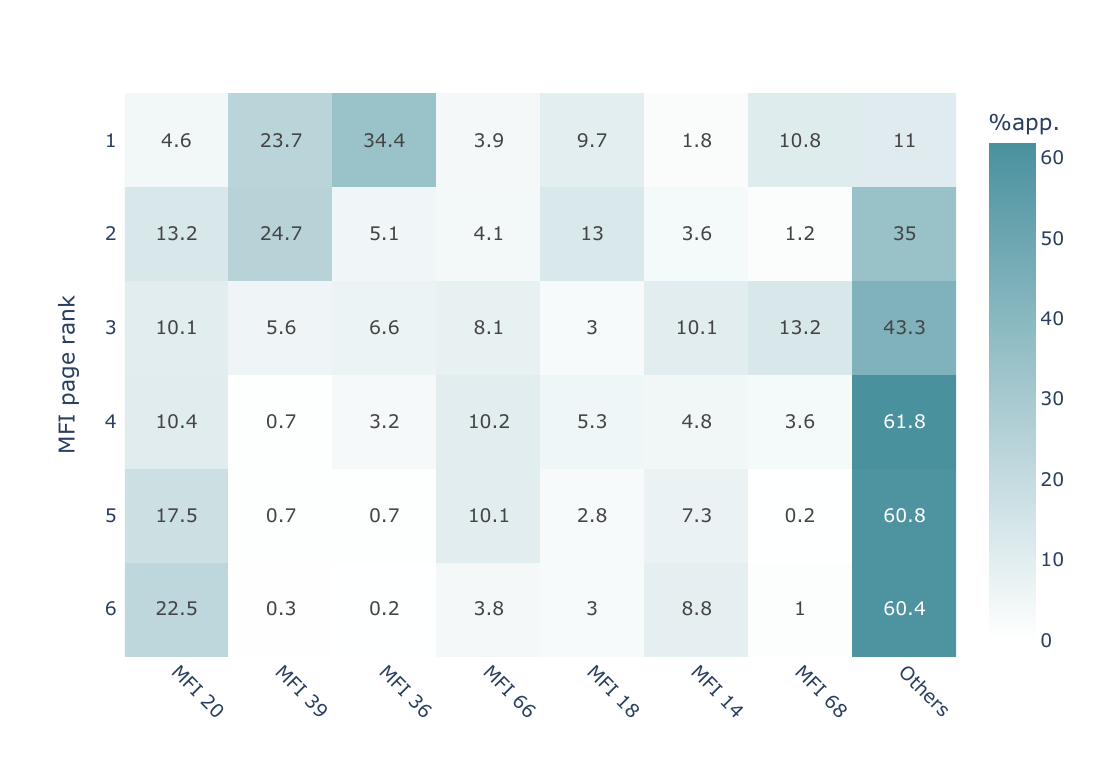}
\centering
\caption{\label{Table_subid5} Percentage of applications by \textit{MFI id} and \textit{MFI page rank}. At different times, and on different web pages, the rank of a fixed MFI may vary, which is the reason why there is no one-to-one correspondence between \textit{MFI id} and \textit{MFI page rank}.  }
\end{figure}

Table~\ref{stat-table} for the five largest MFIs clearly shows that there is an imbalance problem in the data, which is expressed in the dominance of some MFIs: $7\%$ MFIs get more than $27\%$ applications. This can be explained by two factors: first, the psychological tendency of users to trust recommender systems (that is, they are more likely to click on items higher in the ranked list), and second, the popularity of some MFIs themselves. From Fig.~\ref{Number_requests_for_rank} it is clear that MFIs on higher positions receive a larger number of applications, but we can also (somewhat unexpectedly) notice that the line corresponding to MFI with \textit{MFI page rank}  $= 5$ is located higher than those with \textit{MFI page rank}   $= 2$, $3$ or $4$. Which may indicate either the popularity of these specific MFIs or the clients' tendency to somehow choose position $5$  (skipping positions $2$, $3$, and $4$). This information should be taken into account when constructing optimal ranking.
Fig.~\ref{Table_subid5} shows that at certain times there were different MFIs on the fixed \textit{MFI page rank}. This depends on the MFIs ranking and, as was already mentioned, on \textit{page id}. The first position in \textit{MFI page rank} is mainly shared between {``MFI 39''} and {``MFI 36''}, and starting from the second position shares become more dispersed, although there are also some leaders.
This leads us to a conclusion that both \textit{MFI page rank} and \textit{MFI id} influence the number of applications.

\begin{figure}[H]
\includegraphics[height=0.35\textwidth] {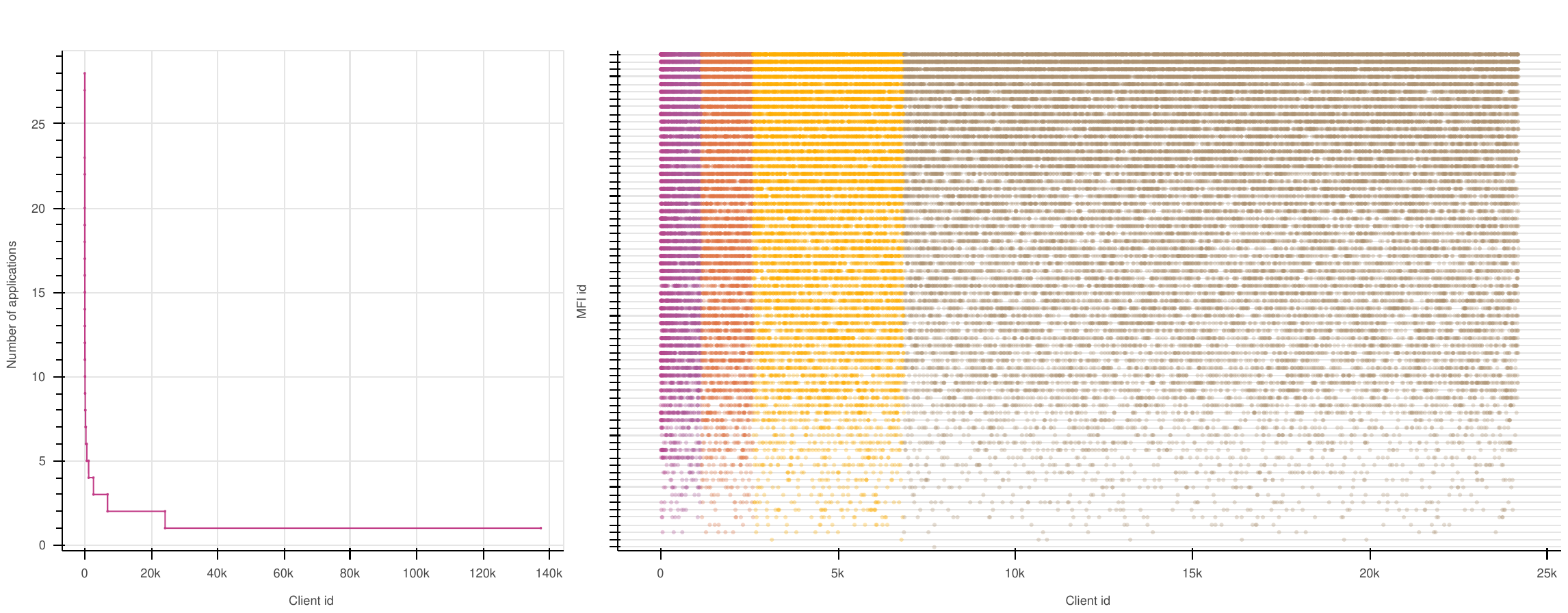}
\centering
\caption{\label{clients_difference} Visualization of clients by their number of applications (left). \textit{MFI id} chosen by clients who submitted more than $1$ application (right). Clients are ordered by decreasing number of applications. Color corresponds to the number of different MFIs for a client. It is easy to see that clients apply to various MFIs.  }
\end{figure}

The problem of imbalance in the number of clients in different MFIs can significantly affect the final ranking. To overcome it, we used Bayesian methods when building the ranking model, which will be described in the Section~\ref{feat_desc}. In the dataset, one can notice that some clients submitted applications to different MFIs. From Fig.~\ref{clients_difference} it is clear that the majority of clients, about $99.5\%$, submitted less than $6$ applications, with $82.4\%$ submitting only one.

\begin{wrapfigure}{r}{0.4\textwidth}
    \centering
\includegraphics[width=0.35\textwidth]{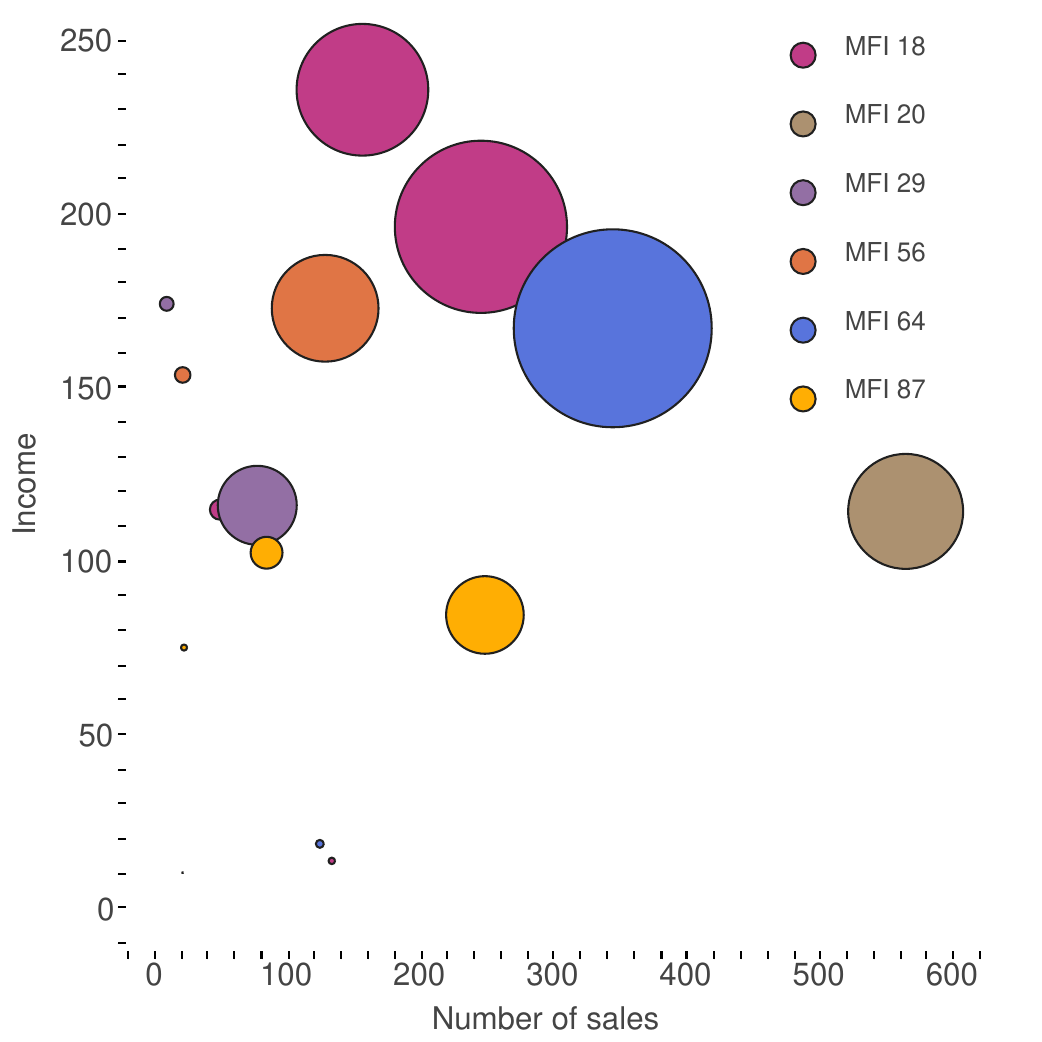}
\caption{\label{income_4_mfi}  Income, number of sales and EPC for $4$ indicative MFIs.}
\end{wrapfigure}

The ``Income, mean'' and ``Income, $\sigma$'' columns in the Table~\ref{stat-table} show that \textit{income} (the fee for an approved application) can vary significantly for a given MFI. It can also vary significantly between different MFIs. 
Fig.~\ref{income_4_mfi} shows for each of the $6$ indicative MFIs all possible values of \textit{income} (OY axis), the number of ``sales'' for the corresponding values of \textit{income} (OX axis) and for each pair (\textit{income}, number of ``sales'') the business metric EPC (earning per click) {in the form of area} of a circle equal to 
$$\dfrac{\text{\textit{income}} \times \text{num. of ``sales''}}{\text{num. of clicks}}.$$ 
\textit{Income} is a dimensionless quantity determined by the internal policy of the AW and MFIs, which is unknown to us. It is possible to assume that the value of this parameter is related to the client's credit history and loan amount. This parameter has a direct impact on profit and therefore should be considered in the ranking.

\begin{figure}[H]
\includegraphics[width=10cm]{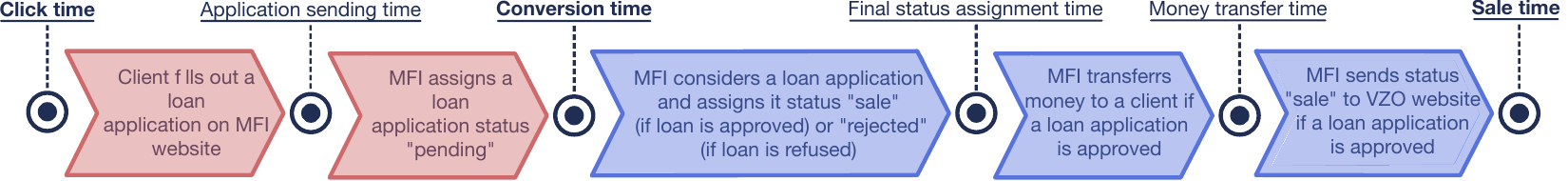}
\centering
\caption{\label{Time_line} Life span in time for a loan. The time periods marked in red and blue we call \textit{conversion period} and \textit{processing period} respectively. }
\end{figure}

In the data the life cycle of an application is marked by three timestamps: \textit{click time}, \textit{conversion time}, and \textit{sale time} if the application is approved. It is important to understand what time intervals are formed by these timestamps. Fig.~\ref{Time_line} shows that, in fact, the life cycle of an application consists of more than three time points. It starts with \textit{click time} when the user on the aggregator website clicks ``Submit an application'' button on the card of the selected MFI and is being redirected to this MFI's website.

Next, the user needs a certain amount of time to fill out and submit an application on the MFI website. When MFI accepts the application, it is assigned \textit{status} ``pending'' and the timestamp is sent to the AW. This timestamp will be recorded in the database as \textit{conversion time}. After this, an MFI employee requires some time to consider the application and make a decision on it. In case of refusal, MFI sends the \textit{status} ``rejected'' to the AW and the application is closed. This time is \textbf{not} recorded in the database. In case of approval, MFI transfers the requested loan to the client within a certain period of time. After payment, MFI sends the \textit{status} ``sale'' to the AW and the time of sending this status is recorded in the database as \textit{sale time}. Note that neither time of considering the application nor the exact time of payment are recorded in the database. 

\begin{figure}[H]
\includegraphics[height=0.35\textwidth]{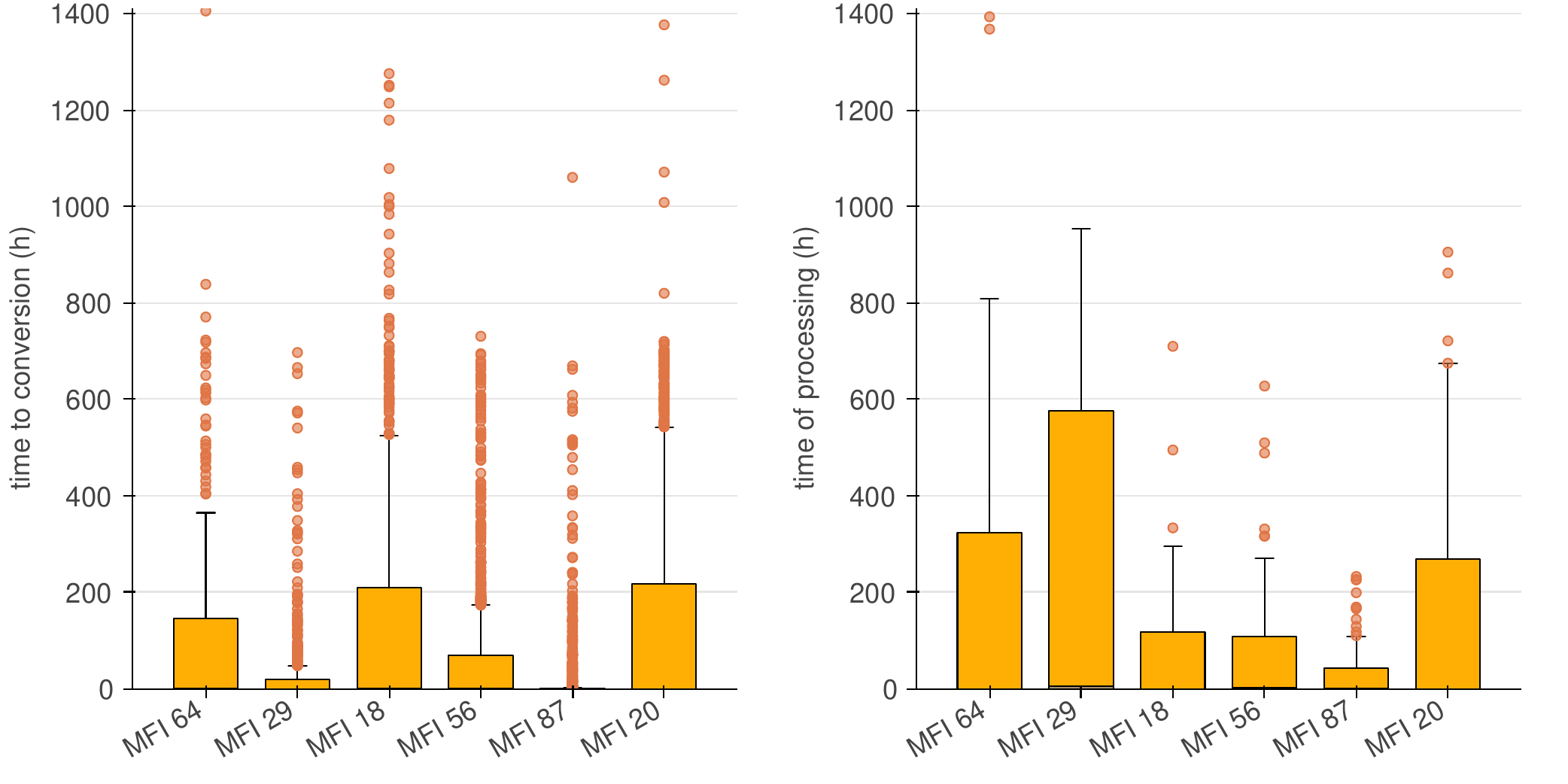}
\centering
\caption{\label{time_6_mfi} Box plots and whiskers for \textit{conversion period} (left) and \textit{processing period} (right) in hours for $6$ indicative MFIs.  Box quantiles: lower $Q_{0.03} $, mid $Q_{0.5}$, upper $Q_{0.97}$; whiskers quantiles: upper $Q_{0.97} + 1.5(Q_{0.97} - Q_{0.03})$, $Q_{0.03} - 1.5(Q_{0.97} - Q_{0.03})$. }
\end{figure}

Due to the presence of unknown time intervals, the processing, analysis and interpretation of temporal data require special attention. Described time periods vary across clients and MFIs (see Fig.~\ref{time_6_mfi}), but mostly are determined by MFIs policy.

\begin{wrapfigure}{l}{0.35\textwidth}
    \centering
\includegraphics[width=0.35\textwidth]{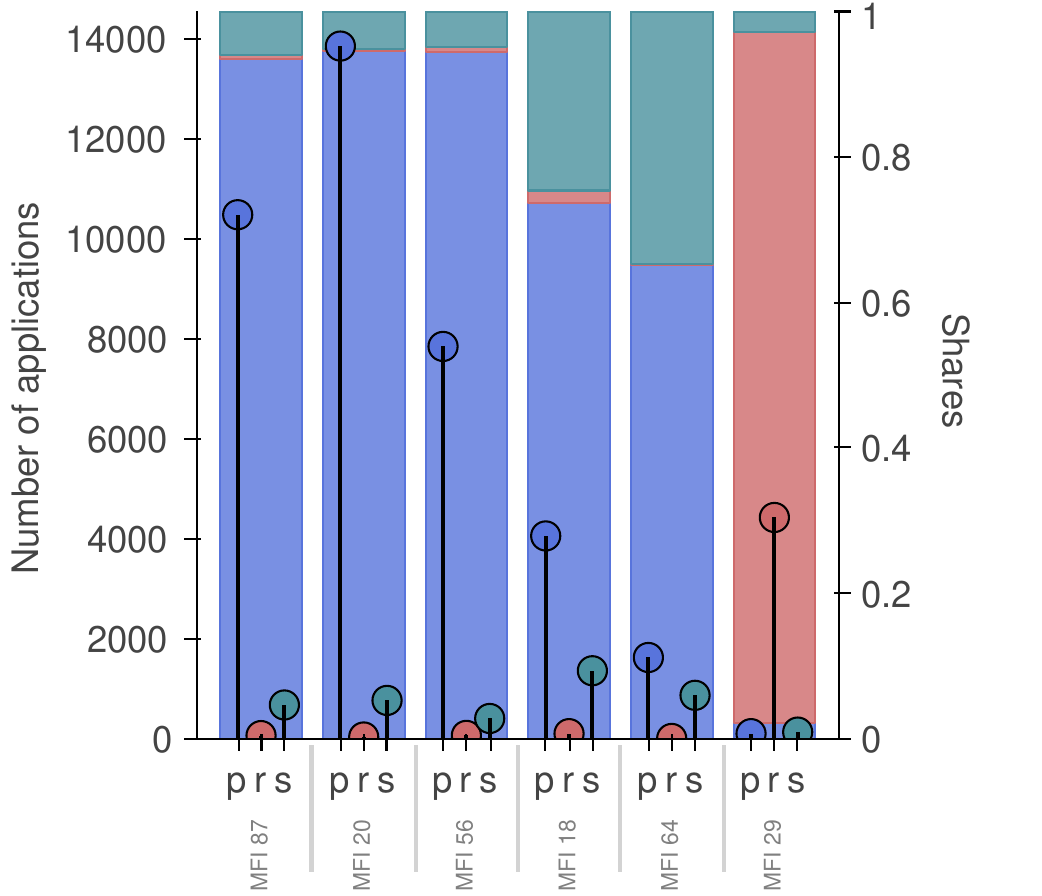}
\caption{\label{status_6_mfi} Number of applications (dots) and shares (bars) for \textit{status} for $6$ indicative MFIs. }
\end{wrapfigure}

Also, given the fact that the final \textit{status} of the application (``sale'' or ``rejected'') is not assigned automatically but sent to the AW by the MFI, the question arises whether MFI is sending the true status. Since the data contains about $77.7\%$ applications with the \textit{status} ``pending'', $13.6\%$ applications with the \textit{status} ``sale'' and $8.6\%$ applications with the \textit{status} ``rejected'', a high percentage of pending applications and a small percentage of refusals raise suspicion. A possible reason for such statistics is that MFIs have no motivation to provide the AW with information when they reject an application (since there is no penalty for not providing it), so even the rejected applications remain with the \textit{status} ``pending''. Only one MFI from the indicative MFIs provided \textit{status} ``rejected'' (see Fig.~\ref{status_6_mfi}).

{\bf Remark about modeling status of an application.} If there were a model for predicting the value of \textit{status} variable (that is, ``sale'' or ``rejected'') for a given application and MFI, we could rank MFIs according to this model as follows: for each application, using a certain set of regressors (for example, \textit{city, OS}, and \textit{loan type}), we would simply select those MFIs that have the highest probability of \textit{status} being ``sale''. Unfortunately, however, the available regressors have a similar impact on the \textit{status} variable for every MFI.

\begin{figure}[H]
\includegraphics[  width= 0.85\textwidth]{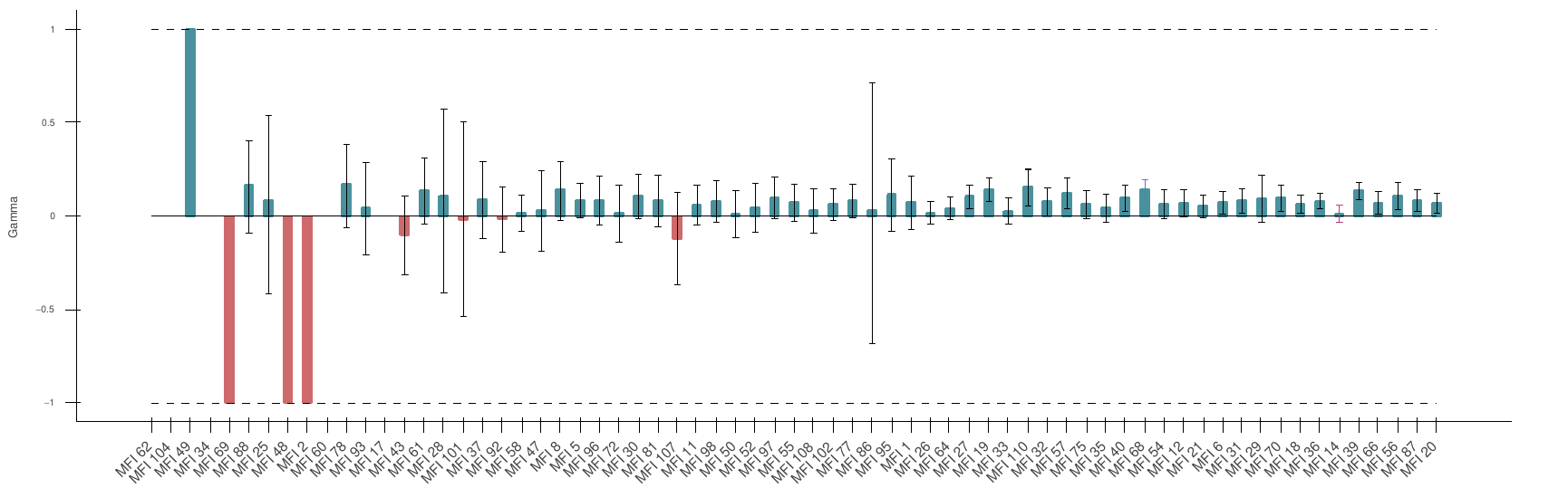}
\centering
\caption{\label{gamma_correlation_ios} 
Сolligation coefficient (bar) with {99.5\%} confidence interval (whiskers) for binary variables: ``sale'' or not, ``iOS'' or not for each MFI. MFIs are ordered according to the increasing number of applications. Drops down correspond to lack of approved applications with ``iOS'' from respective MFIs. Stabilization of the coefficient shows that the dependence between the binary variables is the same among MFIs. Unfortunately, there is no intersection among CIs: a maximum of lower bounds {($0.0919$)} is greater than  minimum of upper bounds {($0.0605$)}.  } 
\end{figure}

 To show that in all MFIs the sale status correlates with the presence of iOS in the same way, we calculated for each MFI  Yule's colligation coefficients $\Upsilon$ between the binary characteristics: ``sale'' or not, ``iOS'' or not (see Fig.~\ref{gamma_correlation_ios}). Note that $\Upsilon$ is analogous to the Spearman's rho for Bernoulli random variables, and uniquely determines the corresponding copula (\cite{Geenens2020}). It is easy to see from Fig.~\ref{gamma_correlation_ios} that the coefficient is about the same for all MFIs with sufficient number of applications. 
A possible explanation for this phenomenon is as follows.  One can assume that the approval rate for users with an ``iOS'' must be higher because their creditworthiness is likely to be higher, but it is naive to assume that for two MFIs with the same approval rate, one of them will approve loan applications submitted by iOS owners significantly more often than the other MFI. After all, if the latter is true, the second MFI must necessarily approve more often applications submitted by clients with different operating systems, i.e. ``less creditworth'' clients.

Inability to create a model using only features given explicitly in the dataset prompted us to construct our own key features and then build a baseline model based on these features.

\FloatBarrier
\subsection{MFIs description dataset}\label{sec_descrip_data}

The attributes of the product dataset are listed in Table~\ref{products-table} in Appendix. For each MFI we again have  \textit{MFI id}, in which regions it operates (\textit{region}) and its operating mode: work schedule (\textit{work schedule}), time when clients can send applications (\textit{application receipt schedule}), and schedule for processing  applications and transferring money to the client if approved (\textit{processing and payment schedule}). Different MFIs have different requirements when applying for a loan: application method (\textit{submission method}), necessary documents (\textit{documents}), and confirmation of a client's identity  (\textit{identification}). Some MFIs can also call potential borrowers, their employers or guarantors (\textit{calls}). Finally, processing of applications consists of: method of processing (\textit{application processing}), time for considering an application (\textit{consideration time}), method of transferring money to the client (\textit{payment method}), speed of transferring money to the client (\textit{payment time}), and method for loan repayment (\textit{repayment method}).

\begin{wrapfigure}{r}{0.35\textwidth} 
    \centering
\includegraphics[width=0.35\textwidth]{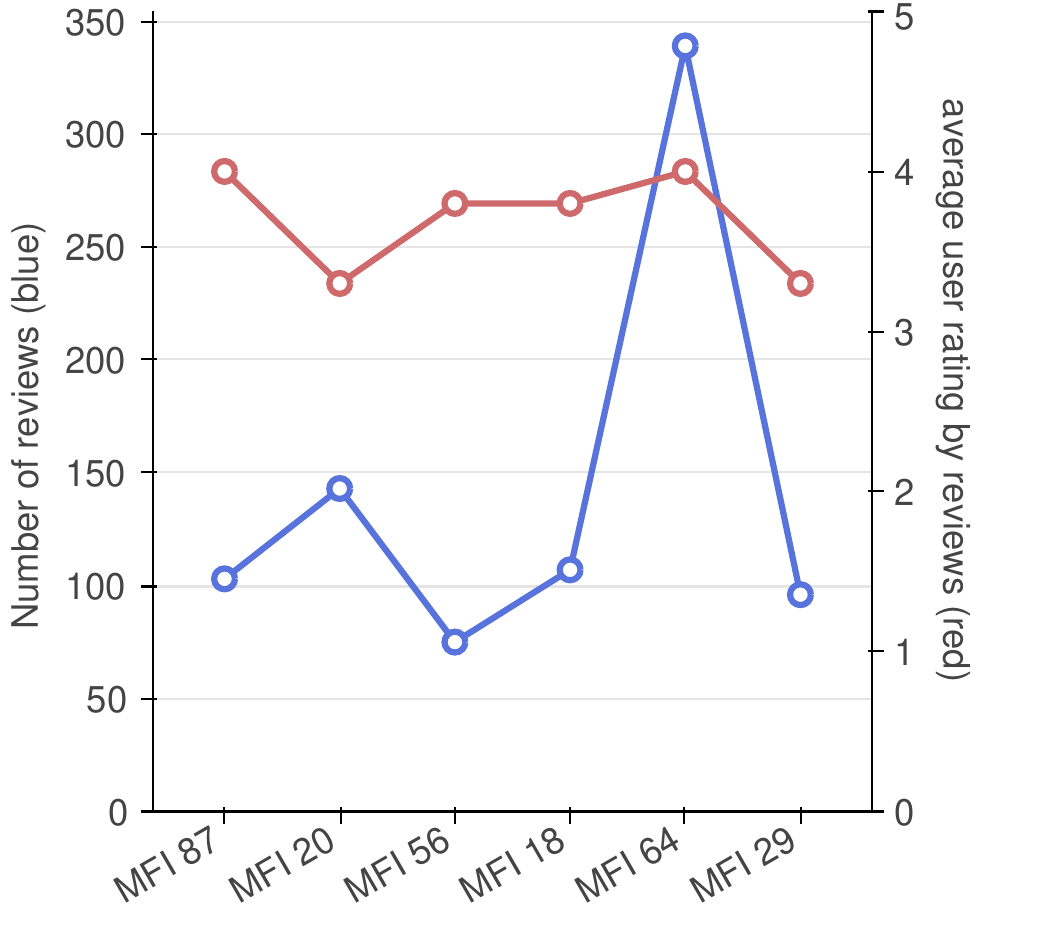}
\caption{ \label{rating_6_mfi} Reviews for 6 indicative MFIs.}
\end{wrapfigure}

The dataset also provides information on clients’ opinions about MFIs: the average rating based on clients' reviews (\textit{average user rating}) and the number of reviews (\textit{number of reviews}) (Fig.~\ref{rating_6_mfi}); whether MFI leaks clients' personal data to the third parties (\textit{unreliability}), whether MFI issues loans to clients with a bad credit history (\textit{bad credit score}) and whether it is possible to extend the loan (\textit {loan extension}).

Due to the fact that MFI can issue loans of three types: standard, long-term and interest-free (\textit{loan type}), each company in the product dataset can be represented by several rows according to the type of a loan. That is, each MFI card is described, and the unique card identifier (\textit{card id}) is consistent with the conversion data (Table \ref{conversion-table}). Terms of the loan include possible amount of money (\textit{loan amount min}, \textit{loan amount max}), repayment period (\textit{loan term min}, \textit{loan term max}), daily interest (\textit{interest min}, \textit{interest max}) and the age of the borrower (\textit{age min}, \textit{age max}).

\subsection{Click dataset}

Click dataset is an extended version of the dataset with conversions (Section~\ref{sec_conversion_dataset}). It includes not only those applications that received a conversion, but also applications that are just clicks out from the AW. We do not know the reason for missing conversions: either the client did not fill out the required form on the MFI's website, or the MFI did not notify us about this application. 

This dataset contains four times more rows than the conversion dataset, so it is published only with the following attributes:  \textit{MFI id, card id, click time, client id, page id,
       MFI page rank, loan type, income}.

The main utility of click dataset is the possibility to calculate business metric earning per click, which has already been seen in Fig~\ref{income_4_mfi}.  

\section{Ranking algorithm}
\label{rank_alg}
To rank different MFIs we first need to compare them. And to do that, it is necessary to determine what features can be used for comparison. These features should reflect the quality of the MFI's work as clearly and as fully as possible. This way we keep the results interpretable; unlike the case when the model itself ``chooses'' the most important features.

We consider the following five key features: average user rating, loan approval rate, fairness, service period and earnings per click. Some of them can be obtained directly from the dataset, while we had to construct the others. 
Our model is based on a pairwise comparison of MFIs according to these features. All numerical values for applications are shown for the ``standard'' {\it loan type}.

\subsection{Feature description} \label{feat_desc}
In this section we present description of the key features, problems we faced when constructing them, and how we solved them. 
Before introducing key features, we need to mention that our ranking algorithm (see Section~\ref{baseline_alg}) does not use their absolute values, since it is based on the \textit{pairwise comparison} of MFIs by each feature. So when creating a feature we are more concerned with ordering by that feature, rather than with the range of its values. Therefore, we also don't use explicit information about how much one MFI is `better' than another.

\begin{figure}[ht]
\includegraphics[height = 4.2cm]{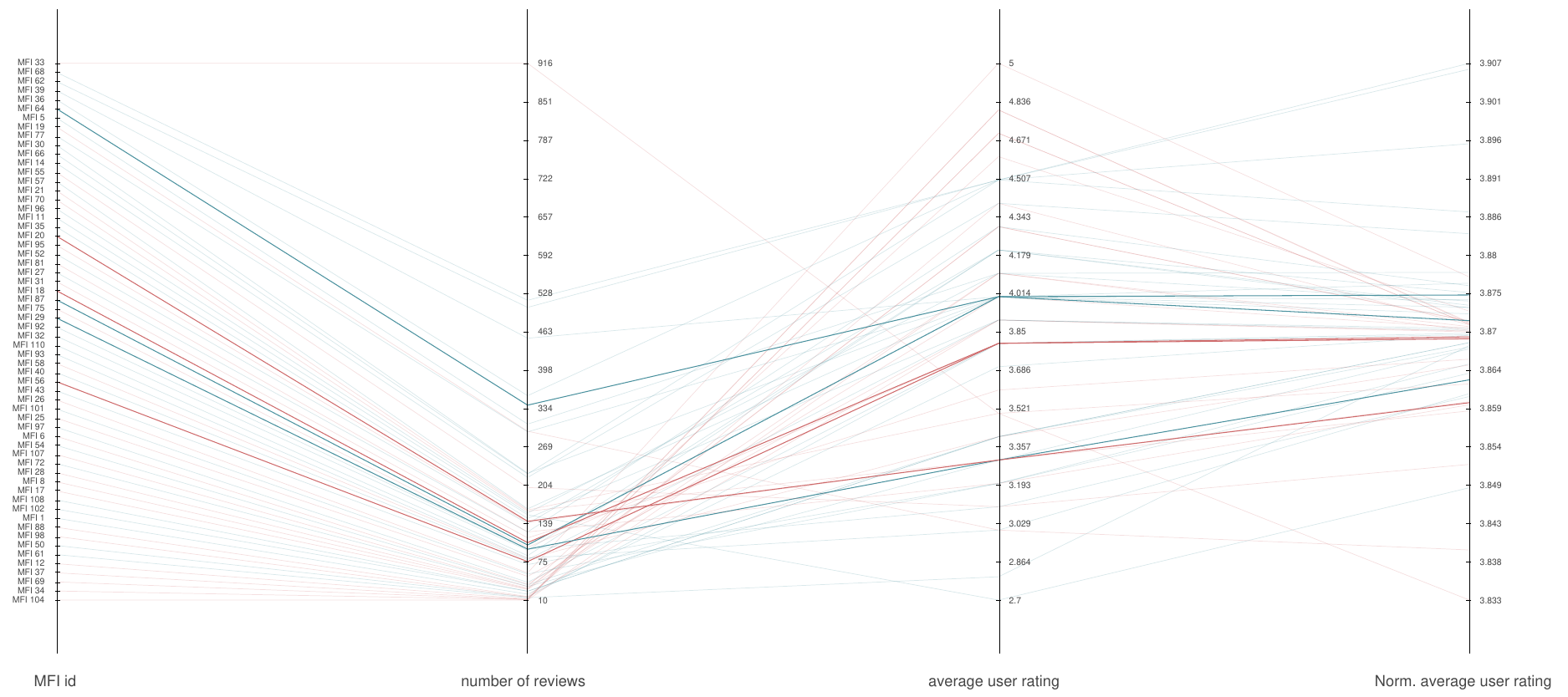}
\centering
\caption{\label{normrate} Parallel plot for \textit{average user rating}  before and after normalization. MFIs whose rank by rating decreased after normalization are shown in red and MFIs whose rank by rating increased --- in green. Indicative $6$ MFIs are highlighted in bold.   MFI $64$ with a significant number of reviews $(339)$ and a positive rating $(4)$ strengthened its position. MFI $20$ with a poor rating $(3.3)$ and a moderate number of reviews $(143)$ lowered its position.  We can also notice that top $4$ MFIs with a great initial rating $(\approx 5)$ lowered its rank as they have very few reviews, but still remained in the top positions.
}
\end{figure}

\textit{{\bf {Average user rating}}} is the only explicit feedback we get from the user, so its utility is obvious. However, we should take into consideration the number of reviews. If the number of reviews is sufficiently large, then average rating is close to the true rating. But when this number is small, the simple average does not adequately describe the rating. For example, an item that has only one review and it's a $5$-star review would have a better average rating than an item that has twenty $4.8$-stars reviews. 
To avoid this inaccuracy we applied Bayesian normalization.
Informally, Bayesian normalization changes user rating in the following way: all initial ratings are divided into two groups, `high' and `low', with `high' rating being greater than or equal to {$3.87$ (weighted average rating)} and `low' rating being less than $3.87$. Then, if several MFIs initially had the same `high' rating, MFIs with more reviews will be ranked higher than MFIs with less reviews. Which is what happened with MFIs $87$ and $64$ with the same `high' rating of $4$ but different number of reviews ($103$ and $339$ respectively), as shown in the Fig.~\ref{normrate}.
Similarly, if several MFIs initially had the same `low' rating, MFIs with more reviews will be ranked lower. 

Technically, we apply Naive Bayes model (\cite{AI20}): given MFI $m$ each review can be treated as independent random variable taking values from $\{1,2,3,4,5\}$ with probabilities
$\{p_{im}\}_{i = 1}^5, p_{im}\geq 0$.  
Probabilities $p_{im}$ are also random variables with prior  Dirichlet distribution 
$Dir(\alpha_i,i=1,\ldots,5),$
where parameters $\alpha_i$ are equal to the total number  of reviews with value $i$ from all MFIs.  
 Posterior predictive distribution for rating of a review  has probabilities
$$\hat{p}_{im} = \frac{\alpha_i + n_{im}}{\sum_i{\alpha_i} + n_m },$$
 where $n_{im}$ is the number of reviews with value $i$ from the $m^{th}$ MFI and $n_m$ -- the total number of reviews  from the $m^{th}$ MFI.
For a normalized \textit{user rating} we use  the posterior mean 
\begin{equation}\label{estimate_rating}
\sum_i i  \frac{\alpha_i + n_{im}}{\sum_i{\alpha_i} + n_m } = \frac{\sum_k n_k \tau_k + n_m\tau_m }{N + n_m }, 
\end{equation}
where $\tau_m$ is the average rating for $m^{th}$ MFI and $N$ is total number of reviews. The equation (\ref{estimate_rating}) allows us to use only average values of ratings, instead of explicit values of each review, that are not available in data. 

\begin{figure}[H]
\includegraphics[width= 0.85 \textwidth]{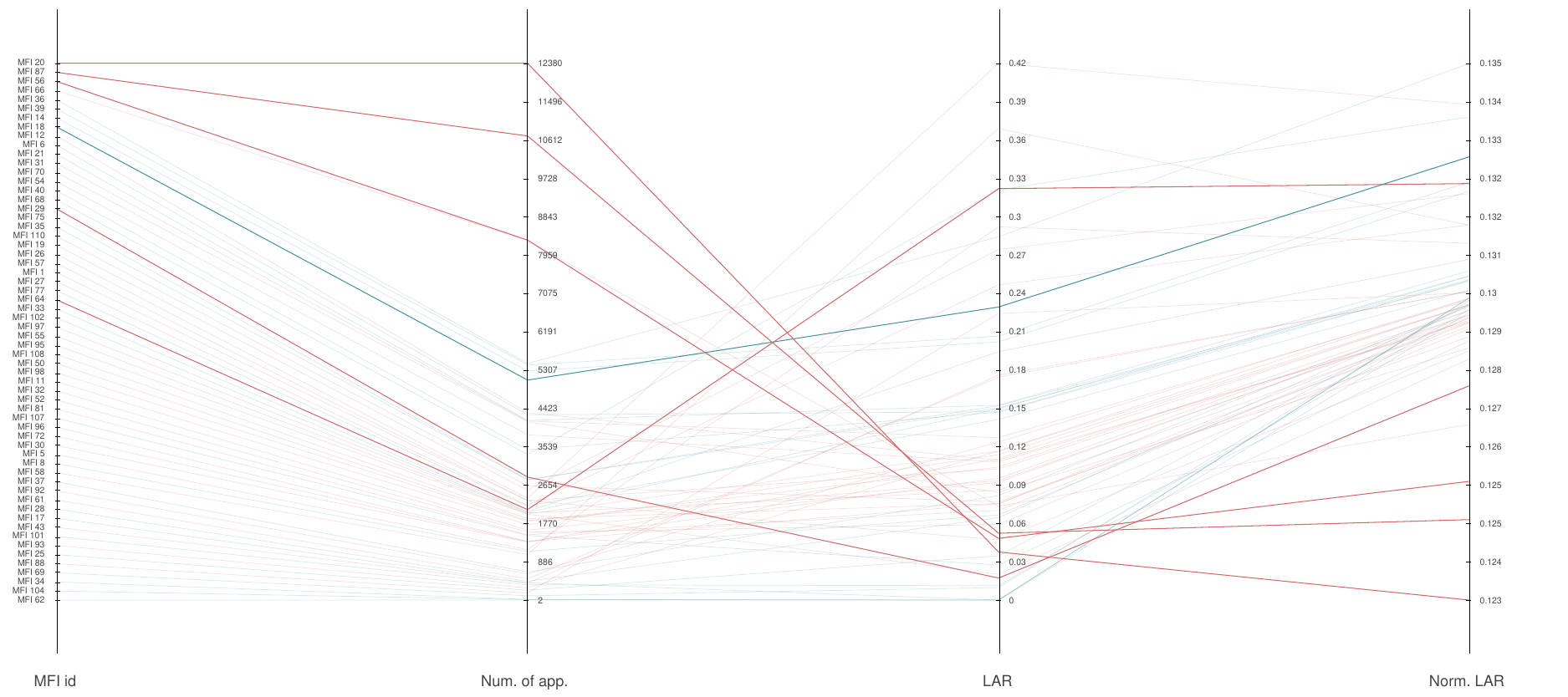}
\centering
\caption{\label{normlar} Parallel plot for \textit{loan approval rate} before and after normalization. MFIs whose rank by LAR decreased after normalization are shown in red, MFIs whose rank by LAR increased --- in green. Indicative $6$ MFIs are highlighted in bold. A large number of applications for MFI $20$, MFI $87$ and MFI $56$  $(12380, 10704, 8302)$ lowered their rank due to weak approval rates $(0.038, 0.052, 0.048)$. On the other hand, a large number of applications for MFI $18$ $(5075)$ significantly strengthened its rather high rank by the approval rate $(0.23)$.}
\end{figure}

\textit{\textbf{Loan approval rate}} (LAR) has the same problem with small number of applications as user rating: we cannot use ratio \(\frac{\text{num. of ``sale''}}{ \text{num. of app.}}\) to adequately describe LAR. So here we again apply Bayesian normalization with Naive Bayes model.
After the normalization we get results (Fig. \ref{normlar}) similar to those for \textit{average user rating}: if initial LAR  is higher than {$0.13$} (average LAR), then the number of application increases normalized LAR. Otherwise normalized LAR decreases as the number of applications increases.  
Also, the normalization is good for handling extreme cases, which, for example, appear at a cold start: if a MFI initial LAR had been equal to $1$, but it had haven only $25$ clients,  the normalized LAR, would significantly had reduced its position relative to other MFIs.

\begin{figure}[H]
\includegraphics[width=12cm]{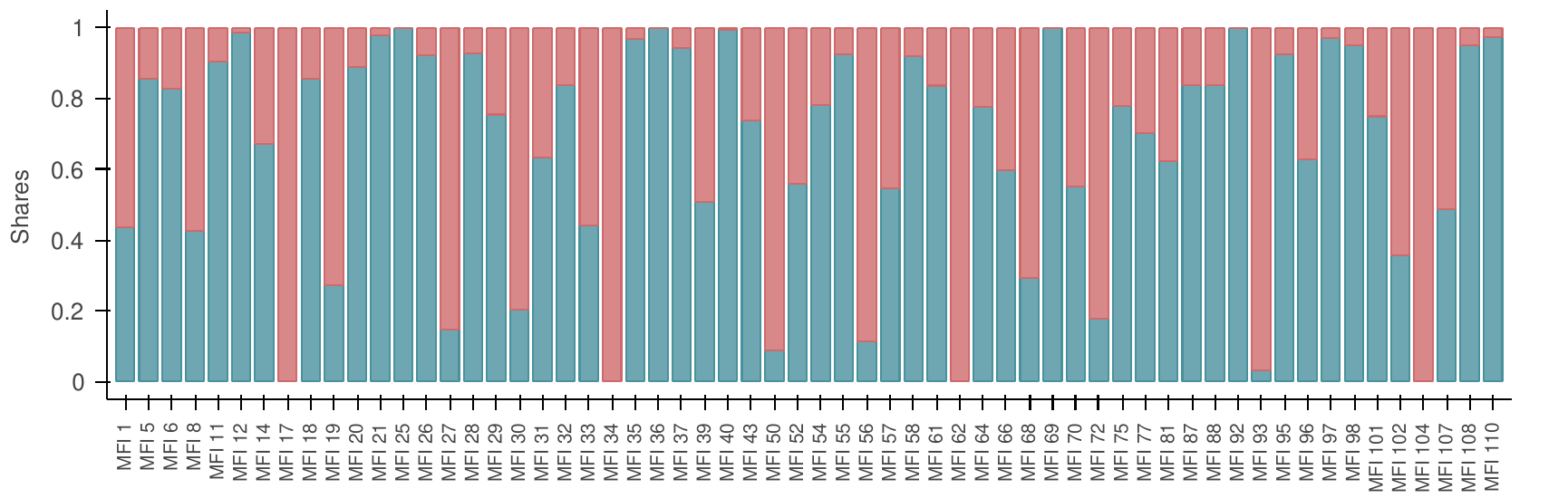}
\centering
\caption{\label{fig_percentage_of_intime_ap} The shares of applications processed within the \textit{consideration time} plus \textit{payment time} are shown in green, while the shares of applications whose processing time exceeds the declared time are shown in red.}
\end{figure}

\textit{\textbf{Fairness}} shows how diligently MFI performs its work: (1) does it provide the correct final status of applications (``sale''/``rejected'') to the AW? (2) Does it change the application status on time? (3) Does it process application on time? (4) Does it leak users' credit score to third parties? 
Positive answer to the first three questions and negative answer to the last one gives MFI one point each and the total value of \textit{fairness} is calculated as the sum of points.
Technically, for each MFI we proceed as follows: MFI gets $+1$ point if
(1) percentage of ``rejected'' applications is higher than $5\%$ and there is at least one ``sale'' in \textit{status}; 
(2) \textit{conversion period} (see Fig.~\ref{Time_line}) of $90\%$ of applications is less then $1$ hour (we will refer to such MFIs as \textit{``on-time'' MFIs});
(3) at least half of the applications are processed according to \textit{consideration time} plus \textit{payment time} from MFIs description data set (see Fig.~\ref{fig_percentage_of_intime_ap});  
(4) value of \textit{unreliability} (see Section~\ref{sec_descrip_data}) is ``False''.

\textit{\textbf{Service period}} shows how long it takes on average from creating a loan application to receiving a final status. The main difficulty with this feature is the correct handling of outliers and filling in missing values. 
Technically, \textit{service period} consists (see Fig.~\ref{Time_line}) of \textit{conversion period} and \textit{processing period}. We transform each of these two periods. 

I. \textit{Conversion period}. Some applications have conversion period of up to a month, which could happen either because user took so long to fill out an application or because MFI took so long to send information about this application to the AW. We believe that if the majority of MFI applications are ``on-time'', then this high value of the conversion period is solely client's fault. 
Thus, we replace all  conversion periods greater than $2$ hours for ``on-time'' MFIs with the median of this MFI's conversion periods.

II. \textit{Processing period}. Unfortunately, in the dataset the processing period is available only for applications with ``sale'' \textit{status}. We fill in missing values (for ``rejected'' and ``pending'' \textit{status}) with the mean of the processing periods for ``sale'' for this MFI. 

\begin{figure}[H]
\includegraphics[width=10cm]{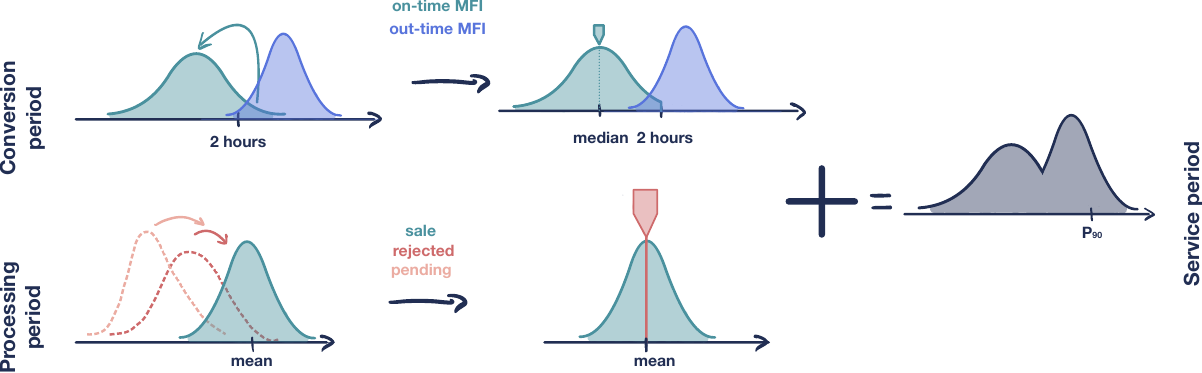}
\centering
\caption{\label{service_period} 
Scheme for obtaining the \textit{service period}. }
\end{figure}

The final \textit{service period} for each application is calculated as the sum of the two aforementioned periods (see Fig.~\ref{service_period}). 
For further comparison between MFIs, we use $P_{90}$ (in seconds): $90^{th}$ percentile of service periods (meaning that lifespan of $90\%$ of applications for a given MFI is less than $P_{90}$ seconds), as this characteristic showed the most adequate results.

\textit{\textbf{Earning per click}} (EPC) directly affects aggregator website's profit so it must be included into our ranking algorithm.  The EPC for MFI $m$ is calculated as the ratio of the total revenue from MFI $m$ to the total number of clicks on MFI $m$
\begin{equation*}
    EPC(m) = \frac{\sum\limits_{\text{``sale'' in } m} \textit{income(``sale'')} }{\text{num. of clicks on } m}.
\end{equation*}

\begin{table}[!ht]
\small
    \centering
    \begin{tabular}{||p{1.3cm}|| p{2.2cm} p{1cm} p{1.1cm} p{2cm} p{1.4cm}||}
    \hline
        \textbf{MFI id} & { \textbf{Average user rating} } & { \textbf{LAR} } & \textbf{Fairness} & \textbf{Service period (sec)} & \textbf{EPC} \\ \hline
        MFI 18 & 3.8687 & 0.1329 & 1 & 126004.2648 & 5.3577 \\ \hline
        MFI 20 & 3.8599 & 0.1229 & 3 & 58758.7124 & 1.5044 \\ \hline
        MFI 29 & 3.8630 & 0.1277 & 4 & 310794.5388 & 1.1219 \\ \hline
        MFI 56 & 3.8690 & 0.1256 & 1 & 68637.8840 & 2.1196 \\ \hline
        MFI 64 & 3.8747 & 0.1323 & 1 & 90024.9961 & 3.5466 \\ \hline
        MFI 87 & 3.8712 & 0.1247 & 3 & 23893.1089 & 1.9553 \\ \hline
    \end{tabular}
    \caption{\label{table_comparison} The key features for $6$ indicative MFIs.}
\end{table}

\subsection{Description of the ranking algorithm} 
\label{baseline_alg}

As was said, our model is based on a pairwise comparison between MFIs according to $5$ key features: \textit{average user rating}, \textit{loan approval rate}, \textit{fairness}, \textit{service period}, and \textit{EPC} (see Section \ref{feat_desc}). There are many ways to compare $5-$dimensional vectors. To ensure that each key feature contributes equally to MFI's final rating, we take into account only the superiority of one MFI over another and  compare MFIs in pairs. First, we set total score of all MFIs to zero. Then, for each key feature we add $+1$ point to the total score of MFI that is `better' of the two based on this feature.  
Note that for \textit{user rating}, \textit{loan approval rate}, \textit{fairness}, and \textit{EPC} `better' means greater and for \textit{service period} 'better' means less.  For example, in Table \ref{table_comparison}, MFI 18 has $3$ points over MFI 20 and, vice versa, MFI 20 has $2$ point over MFI 18.
Thus, we obtain a matrix of pairwise comparisons $\textbf{A} = (a_{ij})$, where the ($i$, $j$) entry shows the number of key features for which the $j^{th}$ MFI outperforms the $i^{th}$ MFI. The matrix for $6$ indicative MFIs is shown in Fig.~\ref{fig_matrix_and_pi}. 

There are many ways in which the comparison matrix $\textbf{A}$ can be used to produce a final ranking list (see, e.g., \cite{Langville2012}). We have chosen a method based on Markov chains because it has helpful interpretability: (step 0) assume that at initial point in time clients are somehow distributed across MFI offices, (step $n\rightarrow n+1$) then each client, based on where the client is, decides which MFI to go to, according to the comparison matrix $\textbf{A}$. The more points new MFI has in comparison with the client's current position, the more likely the client is to move to that MFI. According to ergodic theorem, after a long repetition of steps, the number of clients in each MFI will stabilise regardless of the initial distribution, and will be proportional to the value of the stationary distribution $\pmb{\pi}$ obtained using matrix $\textbf{A}$.

\renewcommand{\kbldelim}{(}
\renewcommand{\kbrdelim}{)}
\begin{figure}[H]
\[
    \underset{\pmb{A}}{
    \vphantom{\begin{pmatrix}0\\0\\0\\0\\0\\0\\0\end{pmatrix}}
    \kbordermatrix{
    & MFI 18 & MFI 20 & MFI 56 & MFI 64 & MFI 66 & MFI 87 \\
    MFI 18 & 0 & 2 & 1 & 2 & 2 & 3 \\
    MFI 20 & 3 & 0 & 3 & 3 & 3 & 4 \\
    MFI 29 & 4 & 2 & 0 & 3 & 4 & 3 \\
    MFI 56 & 2 & 2 & 2 & 0 & 3 & 3 \\
    MFI 64 & 2 & 2 & 1 & 1 & 0 & 2 \\
    MFI 87 & 2 & 0 & 2 & 2 & 3 & 0
  }
  }, \quad
    \underset{\pmb{\pi}^T}{
    \vphantom{\begin{pmatrix}0\\0\\0\\0\\0\\0\\0\end{pmatrix}}
    \begin{pmatrix}
        0.179\\0.129\\0.137\\0.155\\0.199\\0.2\\
    \end{pmatrix}}
\]
\caption{ Comparison matrix and stationary distribution, obtained from Table~\ref{table_comparison}.
}\label{fig_matrix_and_pi}
\end{figure}

Technically, we proceed as follows: find comparison matrix $\textbf{A}$, normalize it so that the sum of each row is equal to $1$, i.e. $\widetilde{a}_{ij} = \frac{a_{ij}}{\sum_{j}a_{ij}}$,  find the stationary distribution by solving the system of linear equations $\pi \widetilde{\textbf{A}} = \pi$, rank MFIs according to probabilities in the stationary distribution $\pi$.  

Next section focuses on evaluating the performance of our ranking algorithm using two comparison methods: A/B testing and our own method based on estimating the probability of loan reapproval.

\section{Comparison methods}\label{sec_comp_method}
To choose the best (i.e., most suitable for our needs) ranking model, we need a way to compare them.  
This is usually done through online or offline methods. In this section we evaluate our ranking algorithm against the current one using two comparison methods: A/B testing and our own method.

The key problems of the comparison of two algorithms are: 
\begin{itemize}
\item in online evaluation, the client cannot visit two MFIs from different ranking lists at the same time;
\item  in offline evaluation, the data is already constructed when some ranking model is running, and it is impossible for all applications to know exactly how they would have been approved by another MFI.  
\end{itemize}

\subsection{A/B testing}

\begin{figure}[ht]
\includegraphics[width= 0.95 \textwidth]{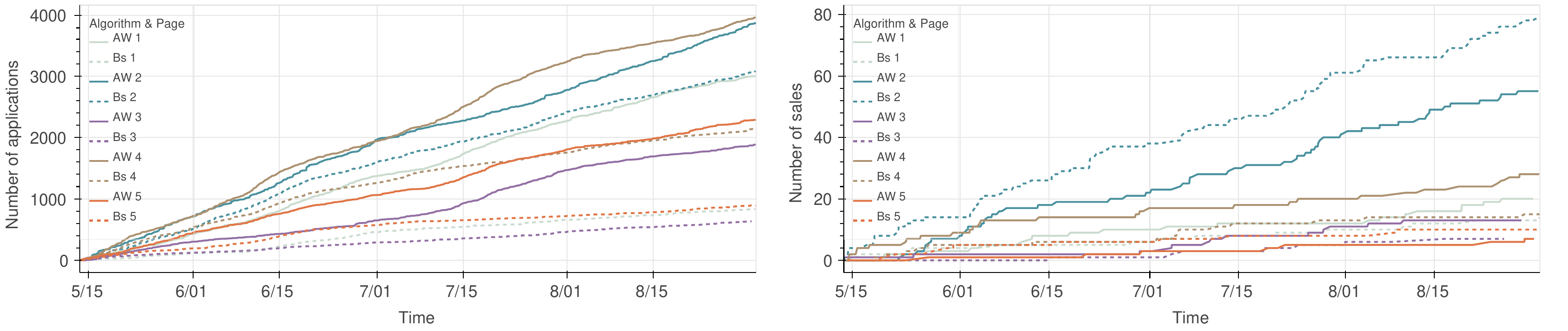}
\centering
\caption{\label{AB_comp_page_clicks_conv} 
The number of applications (left) and the number sales (right) for $5$ \textit{page id}  ranked by AW's algorithm {(solid line)} and by the baseline algorithm {(dotted line)}.
}
\end{figure}

For $5$ \textit{page id} and $3.5$ months we ran A/B testing, where clients are randomly offered either  AW's  ranking or the baseline algorithm. The baseline algorithm was trained at the beginning of each week, AW's algorithm is based on expert evaluation, which is updated manually, updating occurs at least once every two weeks. 
The number of applications and the number sales  by time are illustrated in Fig.~\ref{AB_comp_page_clicks_conv}. As a result of technical reasons, pages were shown not in $50/50$ percent of cases, but in $67/33$. 

\begin{figure}[H]
\includegraphics[width=0.85 \textwidth]{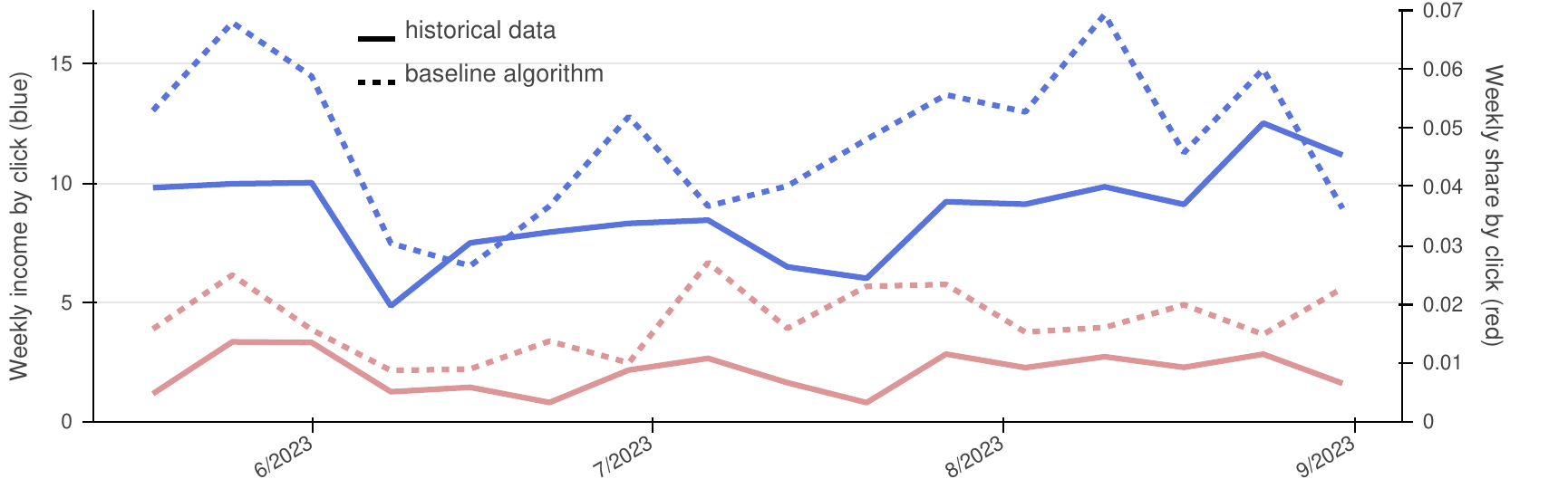}
\centering
\caption{\label{results_with_comparison_method_AB_5page}   
Weekly income  and share by click for the AW and baseline algorithms. It can be seen that the baseline algorithm lines are on average higher than the AW lines.
}
\end{figure}

\begin{figure}[H]
\includegraphics[width=0.85 \textwidth]{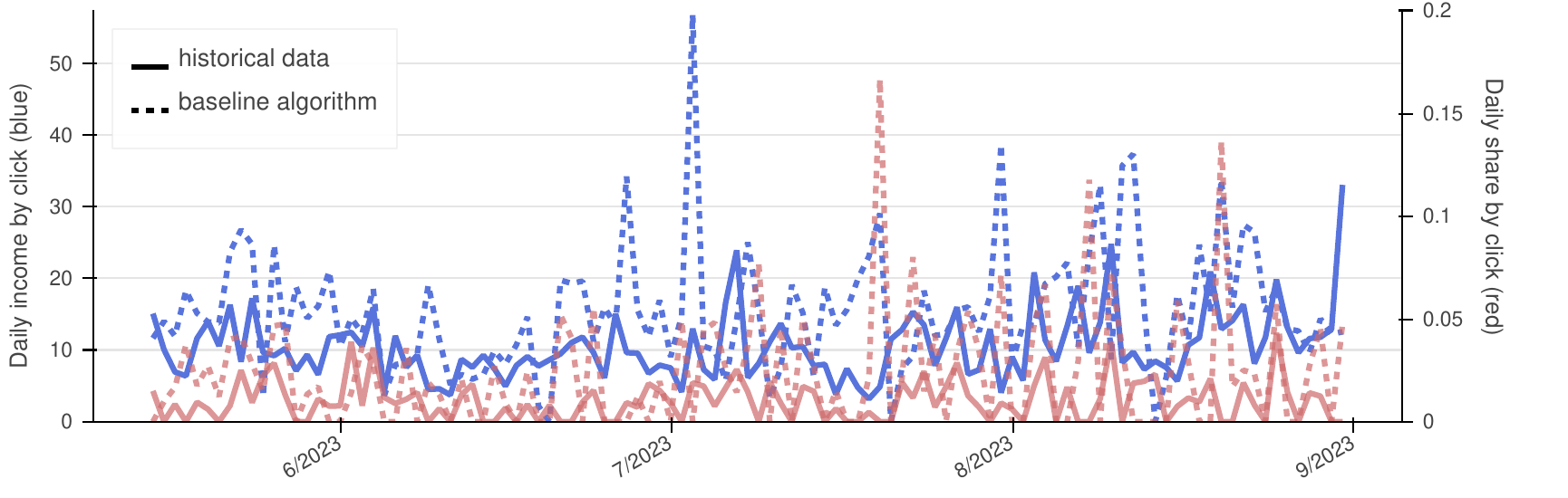}
\centering
\caption{\label{results_with_comparison_method_AB_5page_by_day}   
Daily income  and share by click for the AW and baseline algorithms. It can be seen that the baseline algorithm lines are on average higher than the AW lines.
}
\end{figure}

The baseline algorithm showed  results, which are not worse than the ranking of the AW (see~Fig.~\ref{results_with_comparison_method_AB_5page}, \ref{results_with_comparison_method_AB_5page_by_day}). The A/B testing results, using different key business metrics, are recorded in Table~\ref{table_ab_test}, where in each case the alternative hypothesis was <<baseline algorithm is better than AW's algorithm>>. 

As can be seen from the results for the Fisher test, the main hypothesis that the probability of getting a sale is independent of which algorithm is running on the site is rejected in favour of the alternative that the probability for the baseline algorithm is higher, at the level of $0.05$ in terms of sales per click and sales per conversion. This is due to the higher number of sales per click. For the same reason, the average income per day was higher, but the hypothesis that the average income for the two algorithms is the same is no longer rejected. 
When we examine the average income per sale, the baseline algorithm algorithm displayed unfavorable outcomes. 
However, this metric is poor in this problem, because there is an inadequate algorithm <<show only the highest income MFI>> that shows best results in the metric and bad results overall because of small number of sales.

\begin{table}[!ht]
\footnotesize
    \centering
    \begin{tabular}{||p{2.2cm}| p{2.1cm} | p{2.2cm} | p{2.2cm} | p{1.2cm} ||}
    \hline
        \textbf{Data} & \textbf{test} & \textbf{key statistics (baseline)} & \textbf{key statistics (AW)} & \textbf{$p-$value} \\ \hline
        sales per click rate & Fisher exact test &  $126$ sales,  $7368$ clicks &  $206$ sales,  $14685$ clicks  & $0.045$ \\[2ex]
         \hline
        sales per conv. rate & Fisher exact test &  $126$ sales,  $745$ conv. &  $206$ sales,  $2030$ conv. & $<10^{-4}$ \\[2ex] \hline
        average income per day & Welch's t-test &  average income $1.586$ &  average income $1.433$ & $0.263$ \\[2ex] \hline
         average income per sale  & Welch's t-test &  average income $92.1$ &  average income $112.3$ & $0.997$ \\[2ex] \hline
    \end{tabular}
    \caption{\label{table_ab_test} A/B testing results in key business metrics. The baseline algorithm showed a statistically significant increase in the first two metrics (number of sales per click and number of sales per conversion), a statistically insignificant improvement in average income per day and a decrease in average income per sale.}
\end{table}

\subsection{Comparison method by reapproval}
\label{method_reapproval}

Unlike A/B testing, the method in this section can be applied to the offline data and all \textit{page id}s. That is, the method evaluates how well an algorithm under evaluation (we will refer to it as \textit{verifiable ranking algorithm}, or VRA) works on historical data already generated in some, possibly unknown, way. With this approach, there is no need to use VRA in reality with possible monetary losses.

{\bf The method} relies on predicting whether for a fixed client a loan will be approved by those MFIs the client didn't previously apply to, based on information about approvals by MFIs to which the client applied. 
The essential assumption for the prediction is that the client clicks on the same position in the new ranking list as in the old one. 

The method has a deep technical structure, let us give key steps of the method:  
\begin{itemize}
\item Using all historical loan applications,  compute the probabilities for a loan to be reapproved or redeclined  by each  pair $(i,j)$ of MFIs:
\end{itemize}
$$
\p{MFI_i = \text{``sale''} \,|\, MFI_j = \text{``sale''}}, \ \p{MFI_i = \text{``rejected''} \,|\, MFI_j = \text{``rejected''}}.$$
 
\begin{itemize}
\item Dynamically week by week  generate a ranked list according to VRA. 
\item For each application  calculate the probability of\,           ``sale'' and the mean income as follows:
\begin{equation*}
\begin{split}
& \p{\text{``sale''}}  = \p{MFI_{vra} = \text{``sale''} | MFI_{hist} = x}, \\
& \mathrm{Mean}_{inc}  = \overline{\mathrm{inc}}_{MFI_{vra}}\times \p{\text{``sale''}},
\end{split}
\end{equation*}
 where \textit{hist} is \textit{MFI id} that loan application corresponds to by historical data, $x \in \{\text{``sale''},\text{``rejected''}\}$ is the application \textit{status},
 \textit{vra} is \textit{MFI id} on the new ranked list that has the same place as \textit{hist} MFI had in the `historical' ranked list, $\overline{inc}_{MFI_{vra}}$ is an average income of $MFI_{vra}$ for all of the approved applications.
 \end{itemize}
 
 The equality of ranks  corresponds to the assumption that the client chooses an MFI only on the basis of its place in the ranking list. If a client has already submitted an application to a \textit{vra} MFI, we copy its actual values of \textit{status} and \textit{income}.
 \begin{itemize}
\item  Calculate total \textit{loan approval rate} and total \textit{average income} for all applications for both the historical ranking and VRA. 
\end{itemize}

\begin{figure}[H]
\includegraphics[width=0.85 \textwidth]{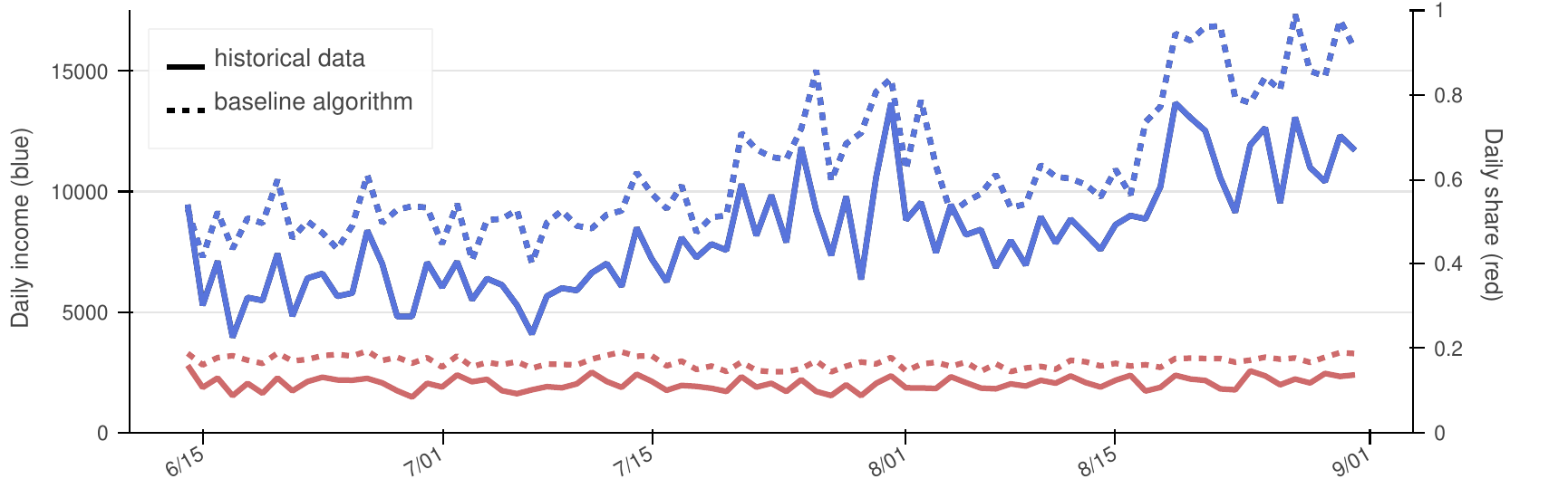}
\centering
\caption{\label{results_with_comparison_method} 
 Daily income and share for historical data and baseline algorithm. The result of comparing the baseline algorithm with historical data, using the comparison method, shows that the loan approval rate almost doubled and  the average revenue increased. }
\end{figure}

{ \bf The results} of the comparison method for the historical ranking and the baseline ranking algorithm  (Section~\ref{rank_alg}) are shown in Fig.~\ref{results_with_comparison_method}, where we took the  \textit{global rank} as the historical rank. Here we use the baseline algorithm with only $3$ key features: average user rating, LAR and EPC, because it shows the best comparison results. We can see that the baseline algorithm  increased the loan approval rate and average revenue.

\section{Conclusion and discussions}\label{sec_discus}
\textbf{Baseline algorithm}. Our goal was to create a ranking model for microfinance institutions in Russia on an aggregated site, which, to our knowledge, has not yet been done. Our algorithm produces a ranked list of MFIs based on features that are important for potential borrowers.

\textbf{Another use of baseline algorithm}. We can use baseline algorithm to measure the quality of new algorithms, for example, those aimed at maximizing profit or approval rate, or algorithms that satisfy a certain set of rules, e.g.: (1) an approved application cannot reduce the MFI rank; (2) a rejected application cannot increase a MFI rating; (3) in order to increase its rank each MFI should employ proactive strategies, like approving an application (as opposed to waiting until competing MFIs drop in the rankings). 

When analyzing data we noticed the presence of change points --- moments of significant changes in MFIs behavior. For example, some MFIs stop sending status ``sale'' for approved  to VZO thus reducing VZO profit. Change point detection can be done automatically and our baseline algorithm allows to understand the influence of change points on ``profit / approval rate''.

One can also study clients dynamic behavior: how (or how often) clients' preferences change over time; and build dynamic models using this information. The contribution of these dynamic models can be estimated using baseline algorithm.   

\textbf{Baseline algorithm problems}. It is noticed that different \textit{page id}s should have  different ranked lists. The simplest way to achieve this is to apply our ranking algorithm for each page separately. But this will lead to a significant reduction in the amount of data. A more robust algorithm is required to rank differently on different pages. 
Furthermore  when performing pairwise comparison, one can take into account not only that one MFI is superior to another but also the degree of this superiority.
    
\textbf{Comparison method}. In addition to the ranking algorithm itself, we also came up with an offline method for evaluating its performance. Our method compares the performance of several algorithms in terms of predicting total profit and total approval rate. This comparison method is based on the probability of an application being reapproved, which is calculated using the frequentist approach. 
    
\textbf{Comparison method problems}. When developing our method we encountered several problems. First, this model requires a huge number of clients, since in order for the estimated probability to be close to the true probability it is desirable to have more than $50$ clients, which for each pair of $68$ MFIs, for each probability gives $ 67*66/2 * 2 * 50 = 221 100$ clients in total. It will take approximately 1 year to reach such number of clients and during this time many MFIs can change their policies or, perhaps, even disappear completely. So, we need a more robust method of estimating the probability of reapproval. 
Second, we assume that when applying for a new loan, the client will click on the same position in the ranked list as previously. Clearly, some clients consider other `properties' of MFIs such as recommendations from people they know personally, MFI advertising and so on. So how do we identify the amount of clients who chose MFI solely due to its rank and the amount of clients who used additional information?
Finally, the comparison algorithm is used in offline settings. An interesting theoretical question arises: ``To what extent can we reduce the impact of the historical ranking model on the comparison result?''

\textbf{Other applications}. It is interesting to see whether our research can be applied to aggregator websites that provide information about microinsurance institutions, where most likely there is also no way to obtain personalized data.

Finally, the publication of the principles the ranking algorithm is based on entails a change in the strategies of MFIs working with the aggregator site. Thus, the ranking model changes not only clients‘ preferences but also MFIs’ behaviour (similar effects arise in recommendation systems theory, e.g. \citet{Adomavicius2013}, \citet{Fleder2009}, \citet{Hazrati2024} show that recommendation systems can have a significant impact on both individual behaviour and the market as a whole).
This raises several interesting theoretical questions, such as: What should the ranking algorithm be like to ensure that the stochastic process of website dynamics reaches a stationary state after its publication? How can we determine the steady state of such a process? Is it possible to identify conditions under which a given stationary process is optimal in some sense?

\subsection*{Data availability statement}
The data that support the findings of this study are openly available in ``Microloans-data'' at https://github.com/TheAuthors/Microloans-data.

\AtNextBibliography{\footnotesize}
\printbibliography

\section*{Appendix}\label{sec_appendix}
\begin{table}[!ht]
\tiny
\begin{center}
\rowcolors{2}{light-gray2}{light-gray}
\begin{tabular}{||lp{5cm}l||}
\hline \textbf{Attribute} & \textbf{Description} & \textbf{Example} \\
\hline MFI id & Unique identifier & ``MFI 1''  \\
loan type & Type of a loan & ``loan-usual'' \\
card id & Unique identifier of MFI card & 1013.0 \\
page id & Page on the aggregator website & ``pensioneram'' \\
MFI page rank & MFI's rank  on the page (at click time) & 4.0\\
MFI global rank & Rank by the aggregator website & 4 \\
click time & When user clicks on the MFI card & 2023-03-31 23:47:00\\
conversion time & When user sends filled application to the MFI & 2023-03-31 23:56:00\\
sale time & When MFI transfers money to user & 2023-03-31 23:58:00\\
status & Status of an application & ``sale''\\
income & Fee that MFI pays the aggregator website & 197.014925\\
client id & Unique identifier of a client & ``hb16ip''\\
country & Client's country & ``Россия''\\
region & Client's region & ``Chelyabinskaya oblast'''\\
city & Client's city & ``Chelyabinsk''\\
device type & -- & ``Мобильный телефон''\\
device & -- & ``iPhone''\\
OS & -- & ``iOS''\\
device browser & -- & ``Mobile Safari''\\
connection type & -- & ``Сотовая связь''\\
device provider & Unique identifier of a device provider & ``DP 1''\\
\hline
\end{tabular}
\caption{\label{conversion-table} Conversion attributes.}
\end{center}
\end{table}
\FloatBarrier

\begin{table}[!ht]
\tiny
\rowcolors{2}{light-gray2}{light-gray}
\begin{tabular}{||p{3cm}p{6cm}p{2cm}||}
\hline \textbf{Attribute} & \textbf{Description} & \textbf{Example} \\
\hline MFI id & Unique identifier & ``MFI 1'' \\
region & In which regions MFI operates & ``Вся Россия'' \\
work schedule & -- & ``круглосуточно'' \\
application receipt schedule & When clients can apply for a loan & ``круглосуточно'' \\
processing and payment schedule & When applications are being considered and money is transferred & ``круглосуточно'' \\
submission method & -- & ``Онлайн'' \\
calls & -- & ``Заемщику'' \\
documents & Necessary documents when applying for a loan & ``Паспорт'' \\
identification & How MFI  confirms client's identity & ``СМС-код'' \\
application processing & -- & ``Автоматически'' \\
consideration time & How long it takes to process an application & ``в течение 20 минут'' \\
payment method & -- & ``карта'' \\
payment time & -- & ``моментально'' \\
repayment method & -- & ``карта'' \\
average user rating & -- & 3.7 (1-5) \\
number of reviews & -- & 307.0 \\
unreliability & Leakage of clients' personal data & ``Да'' \\
bad credit score & Whether loans can be approved for clients with bad credit score & ``Да'' \\
loan extension & -- & ``Есть'' \\
loan type & -- & ``loan-usual'' \\
card id & Unique identifier of MFI card & 1013.0 \\
loan amount, min & -- & 10000 \\
loan amount, max & -- & 30000 \\
repayment period, min & -- & 10.0 \\
repayment period, max & -- & 30.0 \\
interest, min & -- & 0.4 \\
interest, max & -- & 1.0 \\
age, min & -- & 21.0 \\
age, max & -- & 80.0 \\
\hline
\end{tabular}
\caption{\label{products-table}  Product attributes. }
\end{table}
\FloatBarrier

\end{document}